\documentclass{article}
\usepackage{amssymb}
\usepackage{graphicx}
\usepackage{amsfonts}
\usepackage{axodraw}
\usepackage{color}

\newcommand{\Ref}[1]{(\ref{#1})}
\newcommand{\p}{\partial}

\newcommand{\eqa}{\begin{eqnarray}}
\newcommand{\neqa}{\end{eqnarray}}
\newcommand{\equ}{\begin{equation}}
\newcommand{\nequ}{\end{equation}}
\newcommand{\no}{\nonumber\\}
\def\w{\wedge}
\def\la{\langle}
\def\ra{\rangle}

\newcommand{\ket}[1]{|{#1}\ra}
\newcommand{\mean}[1]{\la{#1}\ra}
\newcommand{\6}{$\{6j\}$}

\def\f{\frac}
\def\d{\delta}

\def\lp{\ell_{\rm P}}

\usepackage{bbm}

\newcommand{\SU}{\mathrm{SU}}

\let\epsilon=\varepsilon
\let\eps=\epsilon

\newcommand{\sorgenti}{
\SetScale{0.1}\SetWidth{1.5}
\SetColor{Black}\Line(211,376)(105,46)\Line(105,46)(360,-45)\Line(360,-45)(211,376)\Line(211,376)(479,122)
\Line(479,122)(360,-44)\DashLine(105,46)(479,122){10}\SetWidth{0.5}\GOval(350,242)(37,37)(0){0.882}
\GOval(285,168)(37,37)(0){0.882}\GOval(235,-7)(37,37)(0){0.882}\GOval(429,48)(37,37)(0){0.882}
\GOval(271,79)(37,37)(0){0.882}\GOval(158,207)(37,37)(0){0.882}
\Text(12,18)[lb]{\tiny{\Black{$J_3$}}}\Text(20,-3)[lb]{\tiny{\Black{$J_1$}}}
\Text(23,5)[lb]{\tiny{\Black{$J_5$}}}\Text(39,2)[lb]{\tiny{\Black{$J_6$}}}
\Text(24,14)[lb]{\tiny{\Black{$J_2$}}}\Text(31,22)[lb]{\tiny{\Black{$J_4$}}}
}
\newcommand{\sorgentiYM}{
\SetScale{0.1}\SetWidth{1.5}
\SetColor{Black}\Line(211,376)(105,46)\Line(105,46)(360,-45)\Line(360,-45)(211,376)\Line(211,376)(479,122)
\Line(479,122)(360,-44)\DashLine(105,46)(479,122){10}\SetWidth{0.5}\GOval(350,242)(37,37)(0){0.882}
\GOval(285,168)(37,37)(0){0.882}\GOval(235,-7)(37,37)(0){0.882}\GOval(429,48)(37,37)(0){0.882}
\GOval(271,79)(37,37)(0){0.882}\GOval(158,207)(37,37)(0){0.882}
\Text(12,18)[lb]{\tiny{\Black{$\eta_3$}}}\Text(20,-3)[lb]{\tiny{\Black{$\eta_1$}}}
\Text(23,5)[lb]{\tiny{\Black{$\eta_5$}}}\Text(39,2)[lb]{\tiny{\Black{$\eta_6$}}}
\Text(24,14)[lb]{\tiny{\Black{$\eta_2$}}}\Text(31,22)[lb]{\tiny{\Black{$\eta_4$}}}
}

\newcommand{\seij}{
\SetScale{0.08}\SetWidth{1.5}
\SetColor{Black}\Line(211,376)(105,46)\Line(105,46)(360,-45)\Line(360,-45)(211,376)\Line(211,376)(479,122)
\Line(479,122)(360,-44)\DashLine(105,46)(479,122){10} }

\newcommand{\triple}{\DashLine(241,120)(241,0){10}\DashLine(346,-105)(241,1){10}\DashLine(136,-105)(241,1){10}}

\begin{document}

\title{\bf Coupling gauge theory to spinfoam 3d quantum gravity}
\author{Simone Speziale\footnote{sspeziale@perimeterinstitute.ca} 
\\ [1mm]
\em\small{Perimeter Institute, 31 Caroline St. N, Waterloo, ON N2L 2Y5, Canada.}}
\date{June 11, 2007}

\maketitle
\begin{abstract}
\noindent We construct a spinfoam model for Yang-Mills
theory coupled to quantum gravity in three dimensional riemannian spacetime. 
We define the partition function of the coupled system as a power series in
$g_0{}^2 G$ that can be evaluated order by order using grasping rules and the recoupling theory.
With respect to previous attempts in the literature, this model assigns
the dynamical variables of gravity and Yang-Mills theory to the same simplices of the spinfoam, 
and it thus provides transition amplitudes for the spin network states of the canonical theory.
For SU(2) Yang-Mills theory we show explicitly that the partition function has a
semiclassical limit given by the Regge discretization of the classical Yang-Mills action.
\end{abstract}
\tableofcontents

\section{Introduction}
The large spin limit has proven to be a useful tool to study the semiclassical physics of the 
spinfoam formalism for loop quantum gravity \cite{carlo}. Recent developments
include the right spacetime dependence of the free graviton propagator \cite{grav}. 
On a fixed spinfoam, this limit is related to
the emergence of Regge calculus \cite{Regge}, a discrete approximation to General Relativity (GR).
This result is very desirable, as all physically interesting quantum field theories have useful discrete 
approximations \cite{Giorgione}. In particular, this limit is transparent in the 
Ponzano-Regge (PR) model for three dimensional riemannian quantum gravity \cite{Ponzano}:
the weights of the spinfoam partition function are, in the large spin limit, exponentials of the Regge action.
Similar results are expected in four dimensions \cite{Baez, asympt2}.
If this is the correct way of studying semiclassical physics, then it should be applied not only
to the partition function, but also to expectation values and to the coupling with matter.
In \cite{Hackett} it was indeed shown that expectation values of geometric quantities in the PR model
tend exactly to their classical values in this limit.
In this paper we continue along this line of investigation, and we 
construct a coupled spinfoam model of GR and Yang-Mills (YM) theory in 3d riemannian spacetime
that we show has the right semiclassical limit.

The relevance of the result concerns also the consistence of the proposed coupling of matter fields
to spinfoam quantum gravity. In fact, the literature on the argument is somewhat limited.\footnote{On the
other hand, there are very promising results on the coupling of particles \cite{PR1, PR3}.
See \cite{matterLQG} for the coupling of matter in the canonical formalism, and \cite{mattergft}
in the group field theory approach.}
Models have been constructed specifically for YM fields \cite{altriym}, but they
suffer from the following problem: the discretization procedure at the root 
of the quantization treats the gravitational and YM connections in a different way. 
The advantage of doing so is that the geometric information of
the quantum gravity sector becomes directly available to define the metric dependence of the YM 
action to be quantized. The drawback is that it leads to a spinfoam model 
where the dynamical variables for GR and YM are attached inhomogeneously to the triangulation
discretizing the spacetime manifold, thus the model does not provide transition amplitudes 
for the canonical theory: the data carried by the spin network states of the canonical
theory would not match the data carried by a 2d slice of the spinfoam.

In this paper we show how one can consistently couple YM theory discretizing
both the gravitational and YM holonomies in the same way. 
The key to do so is the use of the generating functional techniques introduced in \cite{Freidel}.
The same procedure has been recently applied to couple fermionic fields to PR \cite{Fairbairn}.
As a consequence of this procedure, the metric dependence of the YM sector is given by a quantum operator
whose action can be evaluated using the recoupling theory for $\SU(2)$.
We study in detail this operator and show that 
the model has the right semiclassical limit but also non trivial quantum corrections.
The resulting model describes spinfoams with labels attached homogeneously, 
thus it provides transition amplitudes for the canonical theory.

\section{The classical theory}

\subsection{Action for GR}
We represent spacetime by a 3d differentiable manifold $M$, and we 
use the triad field as fundamental variable for GR, as this is more suitable to the spinfoam formalism. 
The triad is an $\SU(2)$-valued 1-form $e^I_\mu(x)$ related to the metric by
\equ\label{triad}
g_{\mu\nu}(x)= e^I_\mu(x) \, e^J_\nu(x) \, \delta_{IJ}.
\nequ
Geometrically, the triad is a homomorphism between the tangent bundle $T(M)$
and a principal bundle $P(M, \SU(2))$ whose fibres take values in the $\SU(2)$ 
group.\footnote{In principle, the triad could take values in $SO(3)$,
which is the proper structure group of pure 3d riemannian GR. The group
is extended to $\SU(2)\simeq \mathbb Z_2 \times SO(3)$, in order 
to allow the coupling with fermions. Note also that
while the bundle $P(M, \SU(2))$ is necessarily trivial, there are two isomorphism 
classes of the bundle $P(M, SO(3))$, characterized by their Stiefel-Whitney number, only
one of which is isomorphic to the tangent bundle of $M$. For a discussion
of a possible role of the topology of the bundle, as well as the
constants $\hbar$ and $G$, see \cite{il2d}.}
Over the bundle $P(M, \SU(2))$ we define a connection 1-form $\omega_\mu^{IJ}(x)$,
with curvature $F(\omega)$.

We use the following first order action for GR,
\equ\label{actionGR}
S_{\rm GR}[e_\mu^I(x), \omega_{\mu}^{IJ}(x)] =\frac{1}{8\pi G} \int_M \;{\rm Tr} \;e\wedge F(\omega).
\nequ
The trace Tr is over the algebra indices.
The variation with respect to the connection gives the torsion-free condition, $d_\omega e=0$, 
as an equation of motion. Therefore $\omega$ is the spin connection.
The variation with respect to the triad gives $F(\omega)=0$. The theory is
invariant under diffeomorphisms acting on the (greek letters) spacetime indices, and $\SU(2)$
transformations acting on the (latin letters) algebra indices. Under these transformations,
all solutions of the equations of motion are equivalent, thus the theory has no local degrees of freedom
\cite{Deser}.
Indeed 3d GR is the topological BF theory with $\SU(2)$ as structure group \cite{Witten}. 

We work with dimensionless coordinates, so that $\omega$ and $F$ are dimensionless
and the triad field has dimensions
of a length. We use units $c=1$, but we keep both $\hbar$ and $G$.
The physical dimensions of the gravitational constant $G$
depend on the spacetime dimension $n$. In SI units, we have $[G]={\rm kg}^{-1} {\rm m}^{n-3}$.
In 3d $[G]={\rm kg}^{-1}$ and thus \Ref{actionGR} has the correct physical dimensions of an action,
namely kg$\cdot$m, in units $c=1$.
We define the 3d Planck length as ${\lp}=8 \pi \hbar G$.

\subsection{Actions for YM}\label{sectionYMaction}
Consider YM theory with gauge group ${\cal G}=\SU(N)$. 
We call $A_\mu^a$ the YM connection. The index $a$ is in the algebra $\mathfrak g$ of
$\cal G$. The YM curvature is $F^a_{\mu\nu}=\p_\mu A_\nu^a-
\p_\nu A^a_\mu + f^{abc}A_\mu^bA_\nu^c$, where $f^{abc}$ are the gauge group structure
constants.
Using the metric as the variable for the gravitational field, the Yang-Mills action on 
$n$-dimensional curved space is
\equ\label{action1}
S_{\rm YM}[g_{\mu\nu}, A^a_\mu]=
\frac{1}{4{g_0}^2}\int d^nx \sqrt{g} \, g^{\mu\rho} \, g^{\nu\sigma} \, F^a_{\mu\nu} \, F^a_{\rho\sigma}.
\nequ
Here $g \equiv \det g_{\mu\nu}$. The YM connection is dimensionless, thus
the coupling constant $g_0$ has dimensions of an inverse action times
$(n-4)/2$ length dimensions (in order for the action \Ref{action1} to have the correct physical dimensions
for every spacetime dimension $n$).
In the 3d case, we have $[\hbar^{-1} g_0{}^{-2}] = [\hbar G]$.
Consequentely we introduce a constant with dimensions of a length, 
called ``YM length'',
$\ell_{\rm YM} \equiv \hbar^{-1} g_0{}^{-2}$. 
The dimensionality of the YM coupling constant in 3d is well known, and it is for instance
an indication of the super-renormalizability of this theory.
As it will become clear below, the only relevant constant of the coupled system GR plus YM is the 
adimensional ratio ${\lp}/\ell_{\rm YM}$.

To describe the coupled system with a unique set of variables, we recast this action in terms of the triad field \Ref{triad}.
This can be done using twice the relation
\equ\label{det}
e \equiv \sqrt{g} = {\rm det}\;e^I_\mu(x) = \frac{1}{3!} \, \epsilon_{IJK} \, \epsilon^{\mu\nu\rho} 
\, e^I_\mu(x) \, e^J_\nu(x) \, e^K_\rho(x)
\nequ
to write \Ref{action1} as
\equ\label{action2}
S_{\rm YM}[e_\mu^I(x), A_\mu^a(x)] = \frac{1}{8g_0{}^2}\int_M d^3x \;e^{-1}
\, \epsilon^{\mu\nu\rho} \, \epsilon^{\lambda\sigma\tau} \, e_\mu^I \, e^I_\lambda \, F^a_{\nu\rho} \, F^a_{\sigma\tau}.
\nequ
Notice that no inverse triad $e^\mu_I$ appears in the action \Ref{action2},
at the price of having to deal with the inverse determinant $e^{-1}$.

The definition $T^I_\mu = \delta S_{\rm YM}[e, A]/\delta e_I^\mu$
gives the equations of motion,
$$
F^I_{\mu}(\omega)-\frac{1}{2}e^I_{\mu} F(\omega) = 16 \pi G \, T^I_{\mu},\hspace{0.8cm}
\p_\mu e^I_\nu -\p_\nu e^I_\mu + \omega^I{}_{J\mu} e^J_\nu 
-\omega^I{}_{J\nu} e^J_\mu = 0, \hspace{0.8cm}
D_\mu(A) \left(e F^{\mu\nu}(A)\right)=0.
$$
As it is well known, the presence of matter does not add degrees of freedom to GR \cite{Deser}.
Namely the equations of motion can be solved without local degrees of freedom in the gravitational sector. 
Local degrees of freedom only come from the Yang-Mills sector of the theory.
Geometrically, we are working with a bundle $P(M, \SU(2), {\cal G})$, where the
fibres are independent. This follows from the fact that the YM
field $A_\mu^a$ does not couple to the connection $\omega_\mu^{IJ}$.

The coupled system of gravity and YM theory can be described using the action $S[e, \omega, A]$ 
given by \Ref{actionGR} plus \Ref{action2}, where
\Ref{actionGR} describes gravity alone, while \Ref{action2} describes both the YM field and the
interaction with gravity. To achieve a more symmetric description, we consider an alternative formulation
of YM theory, which makes use of a first order action: YM 
theory can be written as a deformation of BF theory \cite{bfym}
\equ\label{actionBFYM}
\frac{1}{2}\int d^3x\; \eps^{\mu\nu\rho} B^a_\mu F_{\nu\rho}^a(A) - 
\frac{g_0{}^2}{2} \int d^3x \; \sqrt{g} \, g^{\mu\nu} B^a_\mu B_\nu^a.
\nequ
The auxiliary $B$ field has dimensions of an action.
The classical equivalence with \Ref{action1} can be easily checked
by looking at the following equation of motion,
$$
0= \frac{\delta S}{\delta B_\mu^a} = \frac{1}{2}\epsilon^{\mu\nu\rho} F_{\nu\rho}^a
- {g_0{}^2} \sqrt{g} \, B_a^\mu,
$$
and using $\eps^{\mu\rho\sigma} \eps^{\nu\lambda\tau} g_{\mu\nu}= g g^{\rho[\lambda} g^{\sigma\tau]}$.

At the quantum level, the formal equivalence can be straightforwardly verified performing
the gaussian integration over the $B$ field in the partition function for \Ref{actionBFYM}.
On the other hand, the equivalence of the \emph{perturbative} 
quantization of \Ref{actionBFYM} and \Ref{action2} is rather subtle:
particular care has to be taken in the gauge fixing procedure \cite{Accardi, BFYM}.
These are relevant issues that will however not concern us in the following, where we focus only on defining
the full spinfoam partition function and studying its semiclassical limit.

Introducing the Hodge star defined by $*B=\f1{2\sqrt g} \eps^{\mu}{}_{\nu\rho} B_\mu dx^\nu dx^\rho$
the interacting terms can be written as $B\w *B$, and the coupled action
\equ\label{action4}
S[e_\mu^I(x), \omega_\mu^{IJ}(x), B_\mu^a(x), A_\mu^a(x)] = 
\frac{1}{8\pi G}\int_M \;{\rm Tr}\;e\wedge F(\omega) + 
\int_M\; {\rm Tr}\;B\wedge F(A)-
\frac{g_0{}^2}{2} \int_M \;{\rm Tr}\;B\wedge *B.
\nequ
With abuse of notation, Tr denotes the trace over both algebraic indices $I$ and $a$.
As for \Ref{action2}, the metric-dependent interacting term can be written in terms
of triads only using \Ref{det} twice,
\equ\label{BwB}
B\wedge *B = \sqrt{g} \,g^{\mu\nu}\,B_\mu^a\,B_\nu^a = \f12 \,
e^{-1} \, \eps^{\mu\rho\sigma} \, \eps^{\nu\lambda\tau} \,
e_\lambda^I \, e^J_\tau \, e^I_\rho \, e^J_\sigma \, B^a_\mu \, B^a_\nu.
\nequ

This is the action which we use in this paper to study quantum gravity coupled to YM theory:
it has two free pieces, which 
correspond to topological theories, and an interacting piece with local degrees of freedom.

\section{Discretisation}\label{sectionDiscr}
In the following, we are going to quantise the theory
using functional integrals regularized by means of 
a discretisation of the spacetime manifold.\footnote{Notice that a 
lattice discretization of the action \Ref{actionBFYM} on flat spacetime has been
considered in \cite{Conrady}.} 
To do so, we introduce an oriented triangulation $\Delta$. This is an abstract cellular complex
made out of points ${p}$, segments ${s}$, triangles $t$ and tetrahedra {$\tau$},
together with an operation that allows the identification of the $n-1$ dimensional
boundary of an $n$ dimensional object in $\Delta$.
To this abstract complex is associated its dual $\Delta^*$, with a
one-to-one correspondence between $n$-dimensional objects of $\Delta$
and $3-n$ dimensional objects of $\Delta^*$.  The 2-skeleton $\cal J$ of
$\Delta^*$ is a 2-complex made of called vertices $v$, edges $e$ and faces $f$.

The coupled action \Ref{action4} is invariant under diffeomorphisms of $M$. 
Introducing a fixed triangulation breaks this symmetry, thus 
a procedure to restore it will be necessary in the final theory. 
A possible way of doing so is to use the group field theory
formalism \cite{introgft}. This formalism restores the original diffeomorphism invariance by 
producing in general a sum over triangulations weighted by the corresponding spinfoam amplitude
for each fixed triangulation. We postpone the constuction of such a model for further work,
and we focus here on the preliminary step of defining the model on a fixed triangulation.

On the abstract cellular complex $\Delta$, we introduce the algebra variables
$$
X^I_s\in {\mathfrak su}(2), \qquad B^a_s\in {\mathfrak g},
$$
associated to the segments, and the group variables
$$
g_t\in \SU(2), \qquad U_t \in {\cal G},
$$
associated to the triangles. Furthermore, we define the following quantities:
\equ\label{holonomies}
e^{Z_s} \equiv \mathcal{P}\prod_{t\ni s} g_{t}, \qquad
e^{W_{s}} \equiv \mathcal{P}\prod_{t\ni s} U_{t}.
\nequ
Here $\mathcal P$ means that the product over all the triangles
which share the same segment $s$ is oriented. The orientation is induced
from the orientation of $s$.

Consider the classical (discrete) theory defined by the action
\equ\label{discreteAction}
S[X_s^I, g_{t}, B_s^a, U_{t}] = \hbar \sum_s {\rm Tr} \;[X_s \,Z_s]+
\hbar \sum_s {\rm Tr}\;[B_s \,W_s]-
\hbar \,\lambda\, S_{\rm BB}[X_s^I, B_s^a].
\nequ
The interaction term is given by
\equ\label{discreteInt}
\hbar \, \lambda\, S_{\rm BB}[X_s^I, B_s^a] = \f13 \pi {\hbar}^3 {g_0}^2 G \sum_\tau 
\frac{{\cal C}_\tau[X,B]}{{\cal V}_\tau[X]}, 
\nequ
where we introduced the shorthand notations
\equ\label{shorten2}
{\cal C}_\tau[X,B] \equiv  \f1{4}\sum_{p\in\tau} \sum_{s_i, t_i \in p}
\eps^{s_1s_2s_3} \eps^{t_1t_2t_3} X^I_{s_1} X^I_{t_1} X^J_{s_2} X^J_{t_2} B_{s_3}^a B_{t_3}^a,
\nequ
(here $\eps^{s_1s_2s_3}$ is the completely antisymmetric tensor for the three segments
in a given $p$, with the convention $\eps^{123}=1$ for a right-handed triple) and
\equ\label{shorten}
{\cal V}_\tau[X] \equiv  \f14\sum_{p\in\tau}
\f1{3!} \epsilon_{IJK}X_{1}^I X_{2}^J X_{3}^K,
\nequ
where 123 is a right-handed triple of segments for each point $p$.

We kept the constants out of the definition of $S_{\rm BB}$ for a better clarity in later computations.
The coupling constant between gravity and YM is 
\equ
\lambda= \f13 \pi {\hbar}^2 {g_0}^2 G = \f1{24} \f{\lp}{\ell_{\rm YM}}.
\nequ 

The action \Ref{discreteAction} defines a theory on $\Delta$, 
with variables $X_s^I, g_{t}, B_s^a, U_{t}$. In the rest of this section, we will
show that it is an approximation to the continuum theory with action \Ref{action4}.
Recall that the latter is defined on $M$, with variables $e_\mu^I(x), \omega_\mu^{IJ}(x), B_\mu^a(x), A_\mu^a(x)$. 
Consider an embedding $\iota:\Delta \rightarrow M$, which allows us to think of $\Delta$ as a cellular decomposition
of $M$. 
Using the embedding, we have $\ell_s^\mu \sim \int_{s} dx^\mu$.
For each point $p$ in each tetrahedron $\tau$,
we call $\ell^\mu_s(p)$ the three vectors tangent to the three segments $s$ belonging to $p(\tau)$.
We can think of $\ell^\mu_s(p)$ as a 3 by 3 matrix, and we choose coordinates such that
${\rm det}\,\ell^\mu_s(p) = \f1{3!} \eps^{s_1s_2s_3} \eps_{\mu\nu\rho} \ell_{s_1}^\mu \ell_{s_2}^\nu \ell_{s_3}^\rho=1$.
The embedding $\iota$ pushes forward to an embedding for the 
dual triangulation $\Delta^*$, and we analogously have $\ell_e^\mu \sim \int_{e} dx^\mu$ for
each edge in $\Delta^*$. 

We make the following four identifications.
\begin{enumerate}
\item{The triad field:\equ\label{defX}
e_\mu^I(x) \mapsto X_s^I \equiv  
\frac{1}{{\lp}} e_\mu^I(x)\ell_s^\mu \sim \frac{1}{{\lp}} \int_s\,e_\mu^I(x) dx^\mu.
\nequ
For each segment $s$ in $\Delta$,
we clearly have $\ell_s^2 = g_{\mu\nu}\ell^\mu_s \ell^\nu_s = {\lp}^2 X_s^IX_s^I$.
Therefore the variables $X_s$ represent the segment lengths. We will see below
that in the quantum theory they are observables with a discrete spectrum. This is
the sense in which spinfoams describe a quantum geometry, with discrete properties
for the geometrical observables.}
\item{The connection field:
\equ\label{defg}
\omega_\mu^{IJ}(x)\mapsto g_t\equiv e^{\omega_\mu^{IJ}\ell_{e}^\mu} \sim e^{\int_{e} \omega^{IJ}}.
\nequ
On the continuum, 
the connection is geometrically interpreted as an infinitesimal parallel transport. On a 
discrete setting, this should be properly taken over by a discrete minimal parallel transport.
The group element defined above realizes this property.
}
\item{The field $B$:
\equ\label{defB} B_\mu^{a}(x)\mapsto  B_s^a \equiv  
\frac{1}{\hbar} B_\mu^a(x) \ell_s^\mu\sim \frac{1}{\hbar}\int_s B_\mu^a(x) dx^\mu.
\nequ
The geometric interpretation of this auxiliary field is to provide lengths in the fibre space.}
\item{The YM connection:
\equ\label{defUe} A_\mu^{a}(x)\mapsto  U_{t} \equiv 
 e^{A_\mu^{a} \ell_{e}^\mu }\sim e^{\int_{e} A^{a}}.
\nequ
As for the gravitational connection, the YM connection is discretised as a group element
realizing a finite, though minimal, parallel transport along the edges of the dual triangulation.
}\end{enumerate}
We claim that for each configuration of the continuous fields we can find an embedding $\iota$ such
that the difference between \Ref{action4} and \Ref{discreteAction} is arbitrarily small,
if the variables in \Ref{discreteAction} are interpreted as \Ref{defX}--\Ref{defUe}.

To see it, we start 
applying the Stokes' theorem to the connection around a closed face $f$ (for non--abelian groups, 
the theorem holds up to corrections in the area of the face):
$$
\sum_{e\in \p f} \omega_\mu \ell^\mu_{e} \sim \int_{\p f}\omega \approx \int_{f} F(\omega).
$$
As a consequence of the equality above, the definitions \Ref{defg} and \Ref{defUe}
endow the group variables defined in \Ref{holonomies} with the interpretation
of holonomies,
$$
e^{Z_s} \sim e^{\int_f F(\omega)}, \qquad e^{W_s}\sim e^{\int_f F(A)}.
$$
These group elements appear in the first two pieces of \Ref{discreteAction}.
When the embedding is sufficently refined, and the coordinate areas consequently small, we can expand 
the group elements around the algebra, so that \linebreak $Z_s \simeq {\int_f F(\omega)}$ and 
the first two pieces of \Ref{discreteAction} reduce to the two BF terms of \Ref{action4}.

Consider now the interacting term \Ref{discreteInt}. Let us choose a point $p$ in a tetrahedron,
with 123 a right-handed triple of segments.
Using \Ref{defX} and ${\rm det}\,\ell^\mu_s(p) = 1$, we immediately have
\equ
\f1{3!}\eps_{IJK}X_{1}^I X_{2}^J X_{3}^K \sim  
\f{1}{\lp^3} \, \f{e}{{3!}} = \f{1}{\lp^3} \, V_\tau,
\nequ
where in the last step we used the fact that $e$ is the volume of the parallelepiped,
and there are $3!$ tetrahedra in a parallelepiped. 
Symmetrizing the expression above over the four points of $\tau$ we get \Ref{shorten}.
Analogously, using \Ref{defX}, \Ref{defB} and $ {\rm det}\,\ell_s^\mu(p)=1$ we have
\equ
\sum_{s_i\in p(\tau)} \eps^{s_1s_2s_3} X^I_{s_1} X^J_{s_2}B^a_{s_3}
\sim \f{1}{\hbar \, \lp^2} \, \eps^{\mu\rho\sigma} \, e_\rho^I \, e^J_\sigma \, B_\mu^a.
\nequ
Squaring this and then symmetrizing over the four points, we obtain \Ref{shorten2}.
\bigskip

We summarize the discretisation of the dynamical variables in the following box.

\medskip

\framebox{
\begin{minipage}{7cm}
\begin{tabular}{c|c|}
{\bf GR} & {\bf GR} \\
{\bf smooth variable}   				& {\bf discrete variable}	 \\
& \\
triad $e^I_\mu$ 								& $X_s^I\in {\mathfrak su}(2)$  \\
connection $\omega^{IJ}{}_\mu$ 	& $g_{e}\in \SU(2)$ \\
\end{tabular}
\end{minipage}\hspace{1cm}
\begin{minipage}{7cm}
\begin{tabular}{|c|c}
{\bf YM} & {\bf YM} \\
{\bf smooth variable}   				& {\bf discrete variable}	 \\
& \\
auxiliary field $B^a_\mu$									& $B_s^a\in {\mathfrak g}$ \\
connection $A^a_\mu$						&	$U_{e}\in G$ \\
\end{tabular}
\end{minipage}}

\medskip

\noindent 
Let us add two remarks before proceeding with the quantization.
\begin{itemize}
\item We have chosen to discretise both the gravitational and the YM connection
on the dual triangulations. An alternative procedure, already considered in the
literature \cite{altriym}, is to discretise the YM connection on the segments of the
original triangulation $\Delta$, namely as $A_\mu^a \ell_s^\mu$
as opposed to $A_\mu^a \ell_e^\mu$ used in \Ref{defUe}. 
Since the gravity sector provides the lengths $\ell_s$ of $\Delta$, 
this procedure has the advantage that the geometric information for the YM
sector is directly described. The resulting spinfoam model presents however an awkward
feature: the degrees of freedom of YM and GR are attached to different simplices.
This obstructs a clear connection with the 
canonical formalism. The discretization proposed here, on the other hand, leads
to a homogeneous description, as it will become clear below.
\item
The YM holonomy $U_t$ enters only the topological part of YM: the
interaction term does not contain the group elements representing the 
gravitational and YM holonomies, but only the algebra elements $X_s^I$ and
$B_s^a$. This is a consequence of using the action \Ref{action4} to describe the coupled system
that will be useful to define the spinfoam partition function.

\end{itemize}

\section{Spinfoams and quantisation of BF theory}\label{sectionBF}
The quantum theory of \Ref{action2} can be constructed from the spinfoam partition function
\equ\label{Z} Z_{\rm PR} = \prod_{t} \int_{\SU(2)} dg_{t} \; \prod_s
\int_{{\mathfrak su}(2)}dX_s \; e^{{i} \sum_s {\rm Tr} \, [X_s
\,Z_s]}. \nequ This quantity can be evaluated using the harmonic
analysis of $\SU(2)$ (for details, see for instance \cite{carlo, PR1}),
and one obtains \equ\label{ZPR} Z_{\rm PR} = \sum_{j_s} \prod_s {\rm dim}\,{j_s}
\prod_\tau \{6j\}, \nequ where the sum is over all possible
assignments of half-integers $j$ to the segments of $\Delta$. The
half-integers, or spins, label the irreducible representations of
$\SU(2)$. The quantity ${\rm dim}\,j\equiv 2j+1$ is the dimension of the
representation $j$. Finally, a \6 symbol is associated with each
tetrahedron $\tau$ of $\Delta$. The \6 symbol is the key object of the
recoupling theory of $\SU(2)$, and it depends only on the six $j$s
attached to the segments of the tetrahedron. It is defined in
terms of the Wigner 3m symbols (namely -- up to normalization --
Clebsch-Gordan coefficients) as \equ\label{combin6j}
\left\{\begin{array}{ccc} j_1 & j_2 & j_3 \\
j_4 & j_5 & j_6 \end{array} \right\} \equiv 
\left( \begin{array}{ccc} j_1 & j_2 & j_3 \\
m_1 & m_2 & m_3 \end{array} \right)
\left( \begin{array}{ccc} j_1 & j_5 & j_6 \\
m_1 & m_5 & m_6 \end{array} \right)
\left( \begin{array}{ccc} j_4 & j_5 & j_3 \\
m_4 & m_5 & m_3 \end{array} \right)
\left( \begin{array}{ccc} j_4 & j_2 & j_6 \\
m_4 & m_2 & m_6 \end{array} \right),
\nequ
where the sums over the repeated $m_i$'s are understood.
\Ref{ZPR} defines the Ponzano-Regge model for riemannian 3d quantum gravity
in the absence of matter.

To quantize the coupled system, notice that in \Ref{action4} there is also a second BF action, corresponding to the topological term of the YM action \Ref{actionBFYM}. For SU$(2)$ YM theory, the quantization of this action is the same as in the PR model,
only the interpretation of the variables differ. For ${\cal G}=\SU(N)$ 
with $N>2$ the quantization goes along the same lines described above, but the different
algebraic properties of the group influence the final expression. 
In particular in \Ref{combin6j}, one uses a peculiarity of $\SU(2)$, namely
the uniqueness of the Clebsch-Gordan decomposition. The tensor product of two $\SU(2)$
irreps is reducible into a direct sum of irreps, by means of the formula
$j_1\otimes j_2 =|j_1-j_2|\oplus \ldots \oplus j_1+j_2$. The peculiarity here is that
each irrep in the RHS appears only once.\footnote{The uniqueness of the $\SU(2)$ 
decomposition only concernes the 3d case: for spacetime dimension $n>3$ there
are generalized $3n-j$ symbols appearing, which carry additional
$n-3$ quantum numbers. Therefore, in dimension higher than 3, we have intertwiners
also for $\SU(2)$.} This uniqueness is lost for $\SU(N)$ with
$N>2$. Consider for instance $\SU(3)$, and let us use the Cartan highest weight
notation to label the irreps; the tensor product of the
two fundamental representations $(1,0)\otimes(0,1)$, 
contains twice the adjoint irrep $(1,1)$. As a consequence, the six labels on the segments
are not enough to completely characterize a gauge invariant state. The additional numbers required
are called intertwiners, and they are carried by the 3m symbols. In the $\SU(3)$ example above, they distinguish between the two adjoints $(1,1)$, which have the same irrep labels.
Recalling that
combinatorially the 3m symbols are associated to triangles, it should be clear that this additional data 
can be used to label the triangles of $\Delta$. Let us indicate by $\{i_t\}$ a collection
of intertwiners associated with the triangles $\Delta$. Call $\rho$ the
label for the irreps of $\SU(N)$; the partition function \Ref{ZPR}
generalizes for ${\cal G}=\SU(N)$ to
\equ\label{ZBF1}
Z_{\rm BF}[{\cal G}] = \sum_{\{\rho_s, i_t\}} \;\prod_s {\rm dim} \, \rho_s \; 
\prod_\tau \;A_\tau(\rho_s, i_t).
\nequ
Only the labels for the segments and triangles belonging to $\tau$ enter
the tetrahedron amplitude $A_\tau(\rho_s, i_t)$. 
When $N=2$, the tetrahedron amplitude reduces to the $\{6j\}$ symbol described above. In the
abelian case U$(1)$, the tetrahedron amplitude is trivial; calling $n_s$ the irrep
labels, we have only the combinatorial condition $\delta (\sum_{s\in t} n_s)$ for each triangle,
which is the lattice equivalent of the Gauss law.

\subsection{Expectation values}
Geometric observables in the PR model are gauge-invariant functions of the variables $X_s^I$.
To compute their expectation values it is convenient to introduce a generating
functional,
\equ\label{genBF}
Z[J] = \prod_{t} \int_{\SU(2)} dg_{t} \; \prod_s \int_{{\mathfrak su}(2)}dX_s
\; e^{{i} \sum_s {\rm Tr} \, X_s \left(Z_s + J_s \right)}.
\nequ
This can be evaluated as described in \cite{Freidel}, to give\footnote{
To be more precise, the generating functional \Ref{genBF} is evaluated using a slightly
different discretization procedure, in which the $X$-variables are not associated
with the segments (which are dual to faces), but with the \emph{wedges}, 2-surfaces
introduced in \cite{Reisenberger} intersecting dual faces and triangles.
For the applications of \Ref{genBF} in the rest of the paper, the wedge variables would
give the same results as the $X_s$, thus for simplicity here we do not use them.}
\vspace{0.7cm}
\equ\label{gen}
Z[J] = \sum_{\{j_s\}}\; \prod_s {\rm dim}\,{j_s} \;\prod_\tau \left(\prod_{s\in\tau} P(J_s)\right)
\parbox[2cm]{1.8cm}{\fcolorbox{white}{white}{
\begin{picture}(0,0) (10,15) \sorgenti \end{picture}}}.
\nequ
\vspace{0.3cm}

\noindent The quantities entering this expression are defined as follows \cite{Freidel}. $P(J)$ is a function that
relates the Lebesgue measure on ${\mathbbm R}^3$ and the Haar measure on $\SU(2)$. 
Parametrizing as usual the algebra elements as $\vec J = \psi \, \hat u$, with $\hat u$ a unit vector 
on the 2-sphere, we have $|J|= \psi$ and
\equ\label{P}
P(J) = \f2{|J|} \, {\sin\f{|J|}2}.
\nequ
The \6 symbols with source insertions is given by
\vspace{0.5cm}
\eqa\label{6jsorgenti}
\parbox[2cm]{1.8cm}{\fcolorbox{white}{white}{
\begin{picture}(0,0) (10,15) \sorgenti \end{picture}}} &\equiv &
\left( \begin{array}{ccc} j_1 & j_2 & j_3 \\
m_1 & m_2 & m_3 \end{array} \right)
\left( \begin{array}{ccc} j_1 & j_5 & j_6 \\
n_1 & m_5 & m_6 \end{array} \right)
\left( \begin{array}{ccc} j_4 & j_5 & j_3 \\
m_4 & n_5 & n_3 \end{array} \right)
\left( \begin{array}{ccc} j_4 & j_2 & j_6 \\
n_4 & n_2 & n_6 \end{array} \right) \times \no\no &&
D^{(j_1)}_{m_1n_1}(e^{J_1}) D^{(j_2)}_{m_2n_2}(e^{J_2})
D^{(j_3)}_{m_3n_3}(e^{J_3}) D^{(j_4)}_{m_4n_4}(e^{J_4})
D^{(j_5)}_{m_5n_5}(e^{J_5}) D^{(j_6)}_{m_6n_6}(e^{J_6}), \neqa 
where the $D$'s are representation matrices. The $J$'s are
attached to the segments of $\Delta$; they are the sources of the
quantum excitations $j_s$. If $J_s=0$ for all segments, \Ref{6jsorgenti} reduces
to the expression for the $\{6j\}$ symbol given above and $P(0)=1$, thus $Z[J]\Big|_{J=0} \equiv Z_{\rm PR}.$

Using the generating functional the expectation value of a gauge-invariant observable $\Phi[X_s^I]$ can be written as
\equ
\mean{\Phi} = \prod_{t} \int_{\SU(2)} dg_{t} \; \prod_s \int_{{\mathfrak su}(2)}dX_s
\; \Phi[X_s^I] \; e^{{i} \sum_s X_s^I g_s^I} =
\Phi\left[-i\f{\d}{\d J_s^I}\right] Z[J]\Big|_{J=0}.
\nequ
To evaluate expressions of this type we need to know the action
of the algebra derivatives on the measure term \Ref{P} and on the \6 with sources \Ref{6jsorgenti}.
To evaluate the first action, notice that odd derivatives of $P(J)$ always vanish in $J=0$, and
\equ\label{dP}
\f{\p^{(2n)}}{\p J^{I_1} \ldots \p J^{I_{2n}}} P(J)\Big|_{J=0} = \f{(-1)^n}{2^{2n}\,(2n+1)} \, S^{I_1 \ldots I_{2n}}, 
\nequ
where $S^{I_1 \ldots I_{2n}}$ is the fully symmetrized tensor, namely $\d^{IJ}$ for $n=1$, $(\d^{IJ}\d^{KL}+\d^{IK}\d^{JL}+\d^{IL}\d^{JK})/3$ for $n=2$, and so on.
These derivatives only produce constant shifts, independent of the irrep label, and thus negligible
at leading order in the large spin limit. The second action can be evaluated noticing that
acting on a group element in the representation $j$, we have
\equ\label{grasp}
\f\d{\d J^I} D^{(j)}(e^{J})\Big|_{J=0} = -i \, T^{I(j)},
\nequ
where $T^{I(j)}$ is the $I$-th generator in the representation $j$. By inspecting \Ref{6jsorgenti},
we see that a derivative acting on $J_s$ attaches to
the segment $s$ an algebra generator in the irrep $j_s$ labeling the segment.
This action is called ``grasping'', and it is common in the spinfoam literature
(see for instance \cite{DePietri, Freidel, Hackett}).
In particular, to study \Ref{discreteInt} we need the action of the double grasping, which enters twice
\Ref{shorten2}, and the triple grasping, which enters the cubic term \Ref{shorten}. 

Let us introduce the following diagrammatic notation,

\equ\label{double1} -\frac{\d}{\d J_{s_1}^I}\frac{\d}{\d J_{s_2}^I}  \equiv
\parbox{0.7cm}{\begin{picture}(0,0) (15,0)\put(21,15){\small $s_1$}\put(21,-18){\small $s_2$}
\SetScale{0.12}\SetWidth{4}\SetColor{Black} \DashLine(201,100)(201,-100){10} \end{picture}},
\qquad
{\cal V}_\tau\left[-i\f{\d}{\d J}\right] =  \f14\sum_{p\in\tau}\f{i}{3!} \eps_{IJK}\frac{\d}{\d
J_{1}^I}\frac{\d}{\d J_{2}^J} \frac{\d}{\d J_{3}^K} \equiv
\parbox{1.1cm}{\begin{picture}(0,0)(15,0)\SetScale{0.12}\SetWidth{4}\triple\end{picture}}.
\nequ
Notice that to shorten the notation, in the second diagram we included 
the sum, which we recall is over the four points in the tetrahedron, and 123 is a right-handed triple for each point.
Concerning the double grasping, below we will need the cases when $s_1$ and $s_2$ coincide or share a point.
In these two elementary cases the graspings give respectively (see \cite{Hackett})\footnote{
The shift $\f14$ in the double grasping on the same segment was omitted in \cite{Hackett}, where we were
mainly interested in studying the leading order of the semiclassical limit, to which it does not
contribute. However notice that this term authomatically leads to the Ponzano-Regge ansatz
$\ell^2=(j+\f12)^2 \equiv j(j+1)+\f14$. I thank Etera Livine for pointing out to me this additional contribution.}
\equ\label{ele}
\parbox[2cm]{.3cm}{\begin{picture}(0,0) (0,0)
\SetScale{0.12}\SetWidth{4}\Line(0,120)(0,-120)\DashCArc(-12,0)(70,80,-80){10}\put(3,-12){\small{$j$}}\end{picture}}
= \left(C^2(j)+\f14\right) \ \parbox[2cm]{.2cm}{\begin{picture}(0,0) (0,0)
\SetScale{0.12}\SetWidth{4}\Line(0,120)(0,-120)\put(3,-12){\small{$j$}}\end{picture}} \ \ ,
\hspace{2cm}
\parbox[2cm]{.8cm}{\begin{picture}(0,0) (0,0)
\SetScale{0.12}\SetWidth{4}\Line(-90,-80)(0,0)\Line(-90,80)(0,0)\Line(0,0)(140,0)
\DashLine(-52,45)(-52,-45){10}\put(-20,-12){\small{$j_1$}}\put(-20,8){\small{$j_2$}}
\put(10,-9){\small{$j_3$}}\end{picture}}
= \f12\Big[C^2(j_3)-C^2(j_1)-C^2(j_2)\Big] \ \parbox[2cm]{1.5cm}{\begin{picture}(0,0) (-20,0)
\SetScale{0.12}\SetWidth{4}\Line(-90,-80)(0,0)\Line(-90,80)(0,0)\Line(0,0)(140,0)
\put(-20,-12){\small{$j_1$}}\put(-20,8){\small{$j_2$}}
\put(10,-9){\small{$j_3$}}\end{picture}},
\nequ
where $C^2(j)=j(j+1)$ is the $\SU(2)$ Casimir.
We do not report the rather lengthy result for the triple grasping, for which we refer to \cite{Hackett}. 
Notice that the only contribution from \Ref{dP} to these graspings is the shift of $1/4$ to the double grasping on the
same segment.
For the expectation values on a single tetrahedron $\tau$, we introduce also the following notation,
\equ \mean{X_{s_1} X_{s_2}}_\tau  \equiv
\Bigg(
\parbox{0.7cm}{\begin{picture}(0,0) (15,0)\put(21,15){\small $s_1$}\put(21,-18){\small $s_2$}
\SetScale{0.12}\SetWidth{4}\SetColor{Black} \DashLine(201,100)(201,-100){10} \end{picture}}
\Bigg|
\parbox{1.1cm}{\begin{picture}(0,0) (5,14) {\seij} \put(32,-5){\small$\tau$}\end{picture}}\;\Bigg),
\qquad
\mean{{\cal V}[X]}_\tau \equiv
\Bigg(
\parbox{1.1cm}{\begin{picture}(0,0) (15,0)\SetScale{0.12}\SetWidth{4}\SetColor{Black}\triple\end{picture}}
\Bigg|
\parbox{1.1cm}{\begin{picture}(0,0) (5,14) {\seij}\put(32,-5){\small$\tau$}\end{picture}}\;\Bigg).
\nequ

We proceed in the same way for the YM sector, introducing a source $\eta_s^a$ for the $B_s^a$ field
and evaluating the generating functional as before, keeping in mind
that for ${\cal G}=\SU(N)$ the partition function
for BF depends also on additional quantum numbers labeling triangles, the intertwiners.
Defining the $\SU(N)$ equivalent of \Ref{6jsorgenti} (where the intertwiners label the 3m symbols), we have

\equ\label{genBFYM}
Z(\eta) \equiv  \prod_{t} \int_{\cal G} dU_{t} \prod_{s} \int_{\mathfrak g}dB_s\;
e^{i \sum_s \left[ {\rm Tr} B_s W_s + {\rm Tr}B_s \eta_s\right]}
= \sum_{\{\rho_s, i_t\}}\; 
\prod_s \; {\rm dim} \, \rho_s \;\prod_\tau \left(\prod_{s\in\tau} P(\eta_s)\right)
\parbox[2cm]{1.8cm}{\begin{picture}(0,0) (5,15) \sorgentiYM \end{picture}}.
\nequ
To distinguish the GR graspings from the YM ones, we picture the latter with curly lines, such as 
\equ\label{doubleYM} -\frac{\d}{\d \eta_{s_1}^I}\frac{\d}{\d \eta_{s_2}^I}  \equiv
\parbox[2cm]{0.7cm}{\begin{picture}(0,0) (15,0)\put(21,15){\small $s_1$}\put(21,-18){\small $s_2$}
\SetScale{0.12}\SetWidth{4}\SetColor{Black} \Photon(201,100)(201,-100){8}{6} \end{picture}}.
\nequ
As above, we have the elementary graspings
\equ\label{YMgrasps}
\parbox[2cm]{.3cm}{\begin{picture}(0,0) (0,0)
\SetScale{0.12}\SetWidth{4}\Line(0,120)(0,-120)\PhotonArc(-12,0)(70,80,-80){8}{6}\put(3,-12){\small{$\rho$}}\end{picture}}
= \left(C^2(\rho)+\f14\right) \ \parbox[2cm]{.2cm}{\begin{picture}(0,0) (0,0)
\SetScale{0.12}\SetWidth{4}\Line(0,120)(0,-120)\put(3,-12){\small{$\rho$}}\end{picture}} \ \ ,
\hspace{2cm}
\parbox[2cm]{.8cm}{\begin{picture}(0,0) (0,0)
\SetScale{0.12}\SetWidth{4}\Line(-90,-80)(0,0)\Line(-90,80)(0,0)\Line(0,0)(140,0)
\Photon(-52,45)(-52,-45){8}{3}\put(-20,-12){\small{$\rho_1$}}\put(-20,8){\small{$\rho_2$}}
\put(10,-9){\small{$\rho_3$}}\put(0,-9){\small{$i$}}\end{picture}}
= f(\rho_i, i) \ \ \parbox[2cm]{1.5cm}{\begin{picture}(0,0) (-20,0)
\SetScale{0.12}\SetWidth{4}\Line(-90,-80)(0,0)\Line(-90,80)(0,0)\Line(0,0)(140,0)
\put(-20,-12){\small{$\rho_1$}}\put(-20,8){\small{$\rho_2$}}
\put(10,-9){\small{$\rho_3$}}\put(0,-9){\small{$i$}}\end{picture}}.
\nequ
Notice that now the non diagonal grasping depends also on the intertwiner label $i$, and $f$ depends
on the gauge group considered. For ${\cal G}=\SU(2)$ there is no intertwiner and 
$f(\rho)=\f12\big[C^2(\rho_3)-C^2(\rho_1)-C^2(\rho_2)\big]$ as before.

Taking the product of \Ref{gen} and \Ref{genBFYM} we define the generating functional
for the two BF theories,

\medskip
\equ\label{genGRYM}
Z[J, \eta] = \sum_{\{j_s, \rho_s, i_t\}}\; 
\prod_e {\rm dim} \, j_s \; {\rm dim} \, \rho_s \;\prod_\tau
\left(\prod_{s\in\tau} P(J_s) P(\eta_s)\right)
\parbox[2cm]{1.6cm}{\begin{picture}(0,0) (5,15) \sorgenti \end{picture}}
\parbox[2cm]{1.6cm}{\begin{picture}(0,0) (5,15) \sorgentiYM \end{picture}}.
\nequ

Using this generating functional we can compute the expectation values of functions of
$X_s^I$ and $B_s^a$. In particular of \Ref{shorten2} which gives the interaction between GR and YM.
With the double graspings for GR and YM defined above the expectation value of \Ref{shorten2} can be computed
from the action of the following operator,
\equ\label{Cgrasping}
{\cal C}_\tau\left[-i\f{\d}{\d J},-i\f{\d}{\d \eta}\right] = - 
\f1{4}\sum_{p\in\tau}\sum_{s_i, t_i \in p}
\eps^{s_1s_2s_3} \eps^{t_1t_2t_3} \f{\d}{\d J^I_{s_1}} \f{\d}{\d J^I_{t_1}} 
\f{\d}{\d J^J_{s_2}} \f{\d}{\d J^J_{t_2}} \f{\d}{\d \eta_{s_3}^a} \f{\d}{\d \eta_{t_3}^a}
= \parbox{0.9cm}{\begin{picture}(0,0) (18,0)
\SetScale{0.12}\SetWidth{4}\SetColor{Black}
\DashLine(201,120)(201,-120){10}\DashLine(241,120)(241,-120){10}\Photon(301,120)(301,-120){8}{6}
\end{picture}},
\nequ
where the diagrammatic notation implicitly includes the sums, so the lines
represent all the possible graspings within the tetrahedron. 
However the action of this operator on \Ref{genBFYM} is ambiguous, as it contains the product
of two GR graspings: for non abelian groups the graspings
do not commute, so to properly define this operator an ordering prescription is needed. We 
do so in the next section, motivated by recovering the right semiclassical limit.
In section \ref{sectionPert} below we come back to this generating
functional and construct the spinfoam model for the coupled system.

\section{The semiclassical limit}\label{SecSemi}
The spinfoam model that we are going to construct is defined on the fixed triangulation
of spacetime introduced in section \ref{sectionDiscr}.
Therefore we do not expect its semiclassical limit to reproduce the continuum action \Ref{actionGR},
but a suitable discretization of it. For the Ponzano-Regge model \Ref{ZPR} this is well understood:
in the limit when all the spins are large the discrete approximation to GR described by 
Regge calculus emerges. 
In this section we briefly recall the way this happens, and use it to fix the ordering prescription
needed to deal with products of graspings.

\subsection{Regge calculus}

It is an old result~\cite{Regge} that on a triangulation $\Delta$,
made of flat tetrahedra, the curvature is
distributional and concentrated on the segments.
All the geometric information is encoded in 
the segment lengths $\ell_s$. The action can be written in
terms of the segment lengths and their deficit angles $\eps_s$,
\equ
\label{Regge1}
S[\ell_s] = \sum_{s\in \Delta} \,\ell_s\, \eps_s(\ell_s).
\nequ 
The sum here is over all segments in $\Delta$.
The dependence of the deficit angles on the segment lengths can be found as follows.
First, the deficit angles are defined in terms of the dihedral angles as
$\eps_s = 2\pi - \sum_{\tau \ni s} \theta_s^\tau$. Then, the dihedral angles of a tetrahedron
can be expressed in terms of the segments, through the well known formula
\begin{equation}\label{V}
 \sin\theta_s = \f32\f{\ell_s \, V}{A_1 \, A_2},
\hspace{2cm} \parbox[3cm]{4cm}{\includegraphics[width=2.5cm]{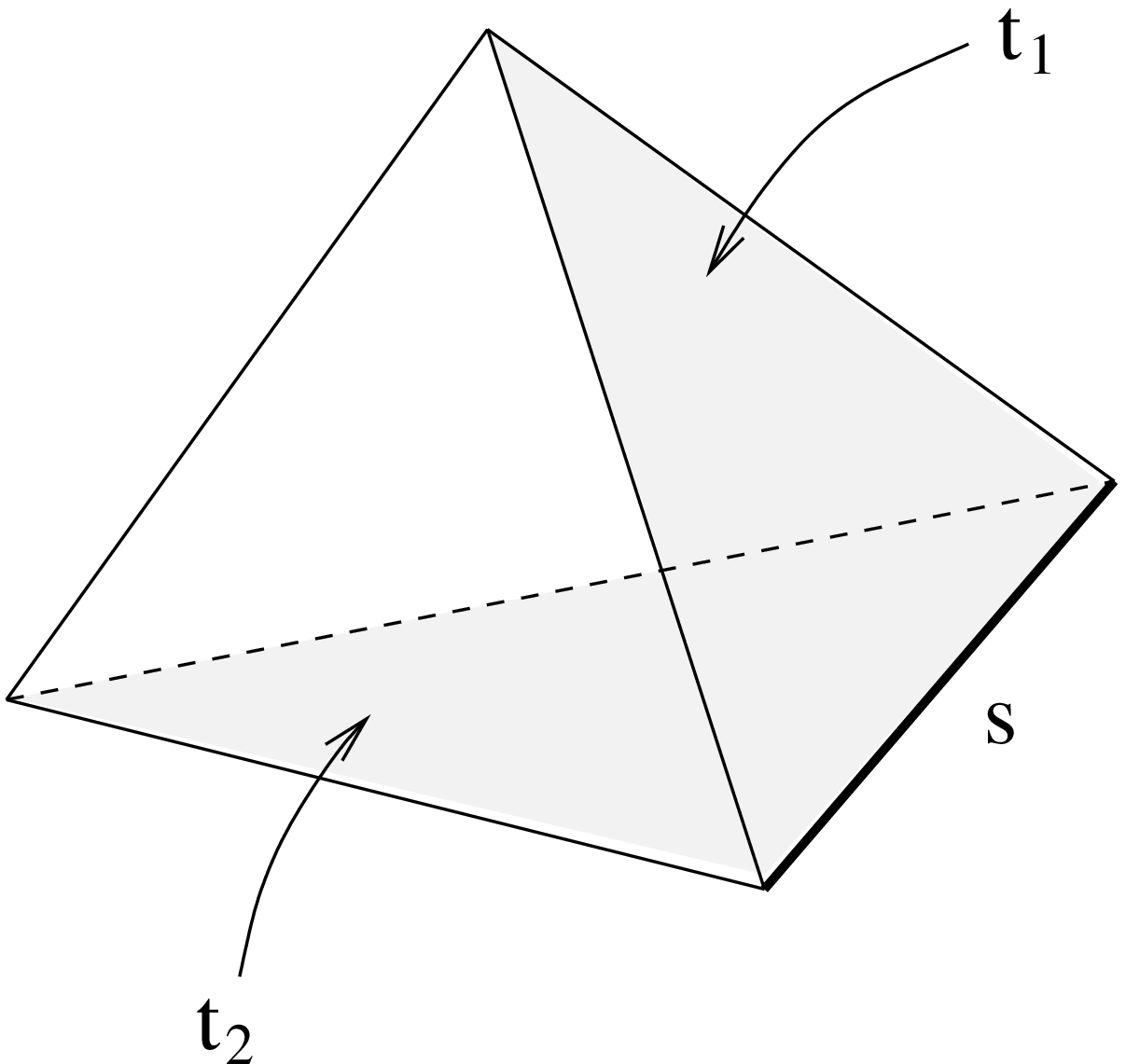}}
\end{equation}
where $A_1$ and $A_2$ are the areas of the two triangles $t_1$ and $t_2$ sharing
the segment $s$, as shown in the above figure. 
Finally the areas $A\equiv V_{(2)}$ and the volumes $V\equiv V_{(3)}$ 
can be expressed in terms of the segment lengths as determinants of the Cayley matrix,
\begin{equation}
\label{cayley}
V_{(n)}^2=\frac{(-1)^{n+1}}{2^n(n!)^2}\det C_{(n)},
\qquad
C_{(n)} =  \left(
\begin{array}{ccccc}
0        & 1         & 1             &   \ldots  &  1             \\
1        & 0         & \ell_1^2      &   \ldots  & \ell_n^2       \\
1        & \ell_1^2  & 0             &   \ldots  & \ell_{2n-1}^2  \\
\ldots   &   \ldots  &   \ldots      &   \ldots  & \ldots    \\
1        & \ell_n^2  & \ell_{2n-1}^2 &   \ldots  & 0
\end{array}
\right),
\end{equation}

A non trivial result of Regge calculus is that the deficit angles are
the conjugate variables to the segment lengths,
\equ\label{deficit3d}
\eps_s = \f{\d S}{\d \ell_s}.
\nequ
Einstein's equations are read from $\delta S / \delta \ell_s =0$ on the bulk.
From \Ref{deficit3d}, we see that they imply a flat triangulation, with zero 
deficit angles everywhere in the bulk. 

To see how this formalism for classical GR emerges from the PR model, we consider
the Regge action on a single tetrahedron,
\equ\label{Regge2}
S_{\rm R}[\ell_s] = \sum_{s\in \tau} \,\ell_s\, \theta_s(\ell_s).
\nequ
The key fact is that exponentials of this action dominate the (homogeneous\footnote{Namely
we rescale the half-integers entering the $\{6j\}$ symbol as $j_s\equiv N k_s$
and take the $N\mapsto\infty$ limit. To keep a simple notation, we do not
write explicitly this rescaling but keep using the variables $j_s$.}) large spin limit of the \6
symbol entering \Ref{ZPR} \cite{Ponzano, asympt2}
\equ\label{asymp} \{6j\} \sim \f{\cos\left(S_{\rm R}(j_s) + \frac{\pi}{4} \right)}{\sqrt{12\pi \, V(j_s)}} , \nequ 
where $V(j_s)$ and $S_{\rm R}(j_s)$ are respectively the classical volume 
\Ref{cayley} and the Regge action \Ref{Regge2}
of the tetrahedron with segment lengths given by $\ell= \lp \left(j+\f12\right)$.
The factor $\f\pi4$ does not change the equations of motion, and the presence of the cosine
as opposed to a single exponential is due to the fact that the PR model sums over both orientations
of the tetrahedron, and the action \Ref{Regge2} is odd under change of orientation
(see discussions in \cite{carlo}).\footnote{This feature will also complicate the analysis of the coupled model
as we discuss below.}

If taking this limit is the correct way to study semiclassical physics, then also the geometric
quantities that one can evaluate in the model should reduce to their classical expressions.
Indeed, in \cite{Hackett} it was shown that the double and triple graspings reduce
to classical scalar products and volumes. For the double grasping on two segments $i$ and $j$
sharing a point, we have the asymptotics
\equ\label{double2}
\Bigg(
\parbox{0.7cm}{\begin{picture}(0,0) (15,0)
\SetScale{0.12}\SetWidth{4}\SetColor{Black} \DashLine(200,120)(200,-120){10}
\end{picture}} \Bigg|
\parbox{1.1cm}{\begin{picture}(0,0) (5,14) {\seij}\end{picture}}\;\Bigg)
 \sim \ell_{s_1}\cdot \ell_{s_2} \f{\cos\big(S_{\rm R}(j_s)+\f\pi4\big)}{\sqrt{12\pi \, V(j_s)}}\, .
\nequ
The scalar product here is defined in terms of the segment lengths in the natural way: if $\ell_{s_3}$
denotes the third segment in the triangle defined by $\ell_{s_1}$ and $\ell_{s_2}$, we have
$2\, \ell_{s_1} \cdot \ell_{s_2} \equiv \ell_{s_3}^2 - \ell_{s_1}^2 -\ell_{s_2}^2$.
For the triple grasping, we have identically zero if the three segments are coplanar, and in any other
case we get
\equ\label{triple2} \Bigg(
\parbox{0.9cm}{\begin{picture}(0,0) (15,0)
\SetScale{0.11}\SetWidth{4}\SetColor{Black}
\put(0,0){\triple}
\end{picture}}
\Bigg|
\parbox{1.1cm}{\begin{picture}(0,0) (5,14) {\seij}\end{picture}}\;\Bigg)
\sim i \, V(j_s) \, \f{\cos\big(S_{\rm R}(j_s)+\f34\pi \big)}{\sqrt{12\pi \, V(j_s)}}. 
\nequ
Notice that the volume is purely imaginary, and it has a different phase in the argument of the cosine.
As described in \cite{Hackett} and anticipated in \cite{Freidel}, this is due
to the fact that the PR model sums over both orientations of spacetime, and
the definition \Ref{shorten} of volume changes sign under
change of orientation.\footnote{This could be avoided introducing a modulus in the
definition of the volume, as is done in the canonical approach \cite{DePietri}.
However, this would prevent a definition of the functional derivatives in terms of graspings.} 
Consequently, real results with the correct phase can be obtained looking at even powers.

As the graspings do not commute
with each other, ordering ambiguites arise in defining the products of graspings. 
As discussed in \cite{Hackett}, we can choose a ``temporal ordering'' $T$ prescribing the graspings to be
performed one after the other. Under this prescription we have
\bigskip

\equ\label{double3}
\Bigg(
\parbox{1.8cm}{\begin{picture}(0,0) (15,0)
\SetScale{0.12}\SetWidth{4}\SetColor{Black} \put(18,-2){$T\{$}\DashLine(280,120)(280,-120){10}
\put(40,0){$\ldots$}\DashLine(480,120)(480,-120){10} \put(60,-2){\}}
\put(31,15){$\overbrace{\hspace{1cm}}^{k}$}\end{picture}} \Bigg|
\parbox{1.1cm}{\begin{picture}(0,0) (5,14) {\seij}\end{picture}}\;\Bigg)
 \sim \left(\ell_{s_1}\cdot \ell_{s_2}\right)^k \, \f{\cos\big(S_{\rm R}(j_s)+\f\pi4\big)}{\sqrt{12\pi \, V(j_s)}},
\nequ

\bigskip

\equ\label{triple3} \Bigg(
\parbox{3.5cm}{\begin{picture}(0,0) (15,0)
\SetScale{0.12}\SetWidth{4}\SetColor{Black} \put(18,-2){$T\{$}
\put(16,0){\triple} \put(60,0){$\ldots$} \put(60,0){\triple} \put(105,-2){\}} 
\put(38,15){$\overbrace{\hspace{2cm}}^{k}$}
\end{picture}}
\Bigg|
\parbox{1.1cm}{\begin{picture}(0,0) (5,14) {\seij}\end{picture}}\;\Bigg)
\sim \f{V(j_s)^k}{\sqrt{12\pi \, V(j_s)}}\, \left\{ 
\begin{array}{lc}
i \, \cos\big(S_{\rm R}(j_s)+\f34\pi\big) & {\rm if} \ k \ {\rm is \ odd}, \\ \\
\cos\big(S_{\rm R}(j_s)+\f\pi4\big) & {\rm if} \ k \ {\rm is \ even}. 
\end{array}\right.
\nequ

This is the way
Regge calculus emerges from spinfoams in the large spin limit: the partition function reduces to exponential
of the Regge action, and the expectation values of the geometry to their classical counterparts.
This means that the theory has a well defined semiclassical limit. Below we are going to construct a spinfoam model
for the coupled system of GR and YM. If this framework is indeed robust, then the coupled system should have 
a semiclassical limit, given by large irrep labels, in which gravity is described
by Regge calculus and YM theory by a discretization \`a la Regge, which we construct below.

\subsection{Regge discretization of YM theory}
We discuss here how \Ref{actionBFYM} can
be written in the Regge approach introduced above. First of all, we consider the topological BF term.
To keep things explicit, let us focus on the special case ${\cal G}=\SU(2)$. By analogy with
\Ref{asymp}, we know that the large $\rho$ limit of \Ref{ZBF1} will be dominated by the cosine
of the discrete action $S=\sum_s {B}_s \, \psi_s(B_s)$ where $B_s=\hbar (\rho_s+\f12)$
represents lengths in the fibre space. The ``dihedral angles''
$\psi_s$ measure the curvature of the fibre space, and the equations of motion impose it flat.
This is just a straighforward discrete version of the BF action.
More interesting is the Regge description of the term \Ref{BwB} which makes YM dynamical
and carries the interaction between GR and YM.
We discretize this by a sum over flat tetrahedra as 
\equ\label{YMaction}
\f{g_0{}^2}2 \int d^3x \, \sqrt{g}\, g^{\mu\nu} \, B_\mu^a \, B_\nu^a \sim
3 g_0{}^2 \sum_\tau \, V_\tau \, g^{\mu\nu} \, B_\mu^a \, B_\nu^a,
\nequ
where we have replaced the canonical volume form $d^3x \, \sqrt{g}$ by 
(six times) the tetrahedron volume $V_\tau$. To express \Ref{YMaction} in
terms of Regge variables, we need the flat inverse metric as a function of the segment lengths.
To obtain this expression, we we start from the formula for the volume $V$ of the tetrahedron,
\begin{equation}
 {(3! \, V)}^2 = g
   =\frac{1}{3!}\,\epsilon^{\mu_1\ldots\mu_n} \, \epsilon^{\nu_1\ldots\nu_n}
    g_{\mu_1\nu_1}\cdots g_{\mu_n\nu_n}.
\end{equation}
Deriving this, we obtain
\begin{equation}
  {(3!)}^2 \, \frac{\p V^2}{\p g_{\mu\nu}}
   = \frac{3}{3!} \, \epsilon^{\mu\mu_2\ldots\mu_n} \, \epsilon^{\nu\nu_2\ldots\nu_n} \, 
      g_{\mu_2\nu_2}\cdots g_{\mu_n\nu_n}= 
      {(3!)}^2 \, {V^2}{ g^{\mu\nu}}.
\end{equation}
Expressing the derivative in terms of the squared edge lengths
$\ell_s^2=g_{\mu\nu}\ell_s^\mu\ell_s^\nu$, we read
\begin{equation}\label{inversemetric}
  g^{\mu\nu} = \frac{2}{V}  \frac{\p V}{\p g_{\mu\nu}}
  = \frac{2}{V} \sum_{s\in\tau}\frac{\p V}{\p\ell_s^2}\frac{\p\ell_s^2}{\p  g_{\mu\nu}}
  = \f2{V}\sum_{s\in\tau}\frac{\p V}{\p\ell_s^2}\ell_s^\mu\ell_s^\nu 
\end{equation}
where the sum is over the six segments in the tetrahedron.

Using \Ref{inversemetric} and \Ref{defB}, we can write \Ref{YMaction} as
\begin{equation}
\label{discreteaction}
  S_{\rm RInt}[\ell_s, B_s] 
  = 3!\, \hbar^2 g_0{}^2 \sum_\tau \sum_{s\in\tau} \frac{\p V_\tau}{\p\ell_s^2} \, B_s^a \, B_s^a.
\end{equation}

To compute the volume derivatives, it is convenient to introduce a double index notation, where a segment
is identified by its vertices, $s\equiv ij$ with $i,j=1\ldots 4$. Let us also introduce
the vectors $\vec n_i$ normal to the triangle obtained removing the point $i$ from the tetrahedron.
We choose their orientation so that they point inward.
These vectors satisfy $|\vec n_i|^2 = 4 A_i^2$, $\vec n_i\cdot\vec n_j = 4 A_i A_j \cos\theta_{ij}$,
where $\theta_{ij}$ are the dihedral angles defined in \Ref{V}.
Then, deriving \Ref{cayley} for $n=3$ and using \Ref{V} we obtain
\begin{equation}
\label{derV}
  \frac{\p V}{\p \ell_{ij}^2}
  = \frac1{72}\, 	\f{\vec n_i\cdot\vec n_j}{V}.
\end{equation}
In conclusion, we can write the interaction term on a Regge triangulation as
\equ\label{ReggeInt}
S_{\rm RInt}[\ell_s, B_s] = \f{1}{12} \, \hbar^2 g_0{}^2 \sum_\tau  \sum_{ij\in \tau} 
\f{\vec n_i\cdot\vec n_j}{V_\tau} \, B_{ij}^a \, B_{ij}^a.
\nequ

We have defined the ordering prescription to deal with the products of graspings, and described
the Regge description of the classical coupled system. 
Below we will show that this is exactly what emerges in the semiclassical limit of the quantum theory.

\section{Spinfoam model of the coupled system}\label{sectionPert}

We now come back to the construction of the spinfoam model for the coupled system.
To define the quantum theory for the coupled system, we evaluate the partition function
\equ\label{Zcoupled}
Z_{\rm GRYM} =  \int {\cal D}e\;{\cal D}\omega\;{\cal D}B\;{\cal D}A\; e^{\frac{i}{{\lp}}
\int {\rm Tr}\,e\wedge F(\omega)  + \frac{i}{\hbar } \int {\rm Tr}\,B\wedge F(A)-
\frac{i}{\hbar }\f{g_0{}^2}2 \int {\rm Tr}\,B \wedge *B}.
\nequ
Firstly, we regularize this formal expression by means of the discretization procedure
discussed above, obtaining
\eqa\label{pert1}
Z_{\rm GRYM} =
\prod_{t} \int_{SU(2)} dg_{t}\prod_{t}  \int_{\cal G} dU_{t} \; \prod_{s} \int_{{\mathfrak su}(2)}dX_s \prod_{s}\int_{\mathfrak g}dB_s \;
e^{i \sum_s \left[ {\rm Tr}X_s Z_s + {\rm Tr} B_s W_s \right]- i \lambda S_{\rm BB}[X, B]},
\neqa
with $\lambda\,S_{\rm BB}$ defined in \Ref{discreteInt}.
We see explicitly from the expression above the advantage of the symmetric form of the action
obtained using the first order formalism for YM theory. In fact, the interaction term
$S_{\rm BB}[X, B]$ is a gauge invariant function of the algebra variables only, and does not
contain gravitational nor YM holonomies. We can  thus
use the generating functional \Ref{genGRYM} and with the substitutions $X_s^I\mapsto -i\frac{\delta}{\delta J_s^I}$,
$B_s^a\mapsto -i\frac{\delta}{\delta \eta_s^a}$, write
$$
Z_{\rm GRYM}= e^{-{i\lambda} S_{\rm BB}[-i\frac{\d}{\d J}, -i\frac{\d}{\d \eta}]} Z[J, \eta]\Big|_{{J=\eta=0}}.
$$
We can interpret the equation above
in the following sense: $Z[J, \eta]$ gives a colouring of the triangulation, 
with labels for $\SU(2)$ and $\cal G$ irreps on the edges, and labels for $\cal G$ intertwiners
on the triangles. 
This colouring provides the kinematical arena for the dynamics of YM and the interaction
between YM theory and gravity, which is realized by the action of the local
operator $e^{{i\lambda} S_{\rm BB}}$. 

To define $S_{\rm BB}[-i\frac{\d}{\d J}, -i\frac{\d}{\d \eta}]$, we use the $T$-ordering
introduced above, but we also need a prescription for the inverse volume $({\cal V}[X])^{-1}$
appearing in \Ref{discreteInt}. 
Motivated by recovering the right semiclassical limit, we define this operator as
\equ\label{invV}
\left({\cal V}_\tau\left[-i\frac{\d}{\d J}\right]\right)^{-1}\equiv 
 \int_0^\infty dt \sum_{k=0}^\infty \frac{(-t)^k}{k!}\, 
 \left({\cal V}_\tau\left[-i\frac{\d}{\d J}\right]\right)^{2k+1},
\nequ
where $t$ is an auxiliary variable with no physical meaning,\footnote{If we compare it
with the Turaev-Viro model \cite{Turaev}, we see that $t$ behaves like the contribution, from
a single tetrahedron, to an imaginary cosmological constant, $\Lambda=it$.} and
the powers of the triple grasping are $T$-ordered as in \Ref{triple3}. 
The advantage of this definition is that, using \Ref{triple3}, we have in the large spin limit
the desired semiclassical behaviour of the inverse volume,
\equ\label{Vmen1}
\mean{\left({\cal V}_\tau\left[-i\frac{\d}{\d J}\right]\right)^{-1}}
\sim i \, \f1{V(j_s)} \f{\cos\left(S_{\rm R}(j_s) + \f34\pi\right)}{\sqrt{12\pi V(j_s)}}.
\nequ

With these definitions, we can evaluate the partition function as a power series in $\lambda$:
\equ\label{pert2}
Z_{\rm GRYM}\equiv  \sum_n \frac{(-i \lambda)^n}{n!}
\left(S_{\rm BB}\left[-i\frac{\d}{\d J}, -i\frac{\d}{\d \eta}\right]\right)^n\, Z[J, \eta]\Big|_{{J=\eta=0}},
\nequ
where
$$
S_{\rm BB}\left[-i\frac{\d}{\d J}, -i\frac{\d}{\d \eta}\right]=
\sum_{\tau} \;\int_0^\infty dt \,\sum_{k=0}^\infty \frac{(-t)^k}{k!} \,  
 \left(V_\tau\left[-i\frac{\d}{\d J}\right]\right)^{2k+1}
{\cal C}_\tau \left[-i\frac{\d}{\d J}, -i \frac{\d}{\d \eta}\right].
$$
The expression \Ref{pert2} provides us with a formula for the partition function \Ref{Zcoupled}
in the spinfoam formalism. It is well defined, order by order in $\lambda$.
The ordering prescription for the operators is as in \Ref{double3} and \Ref{triple3}.
$S_{\rm BB}$ is local because the graspings only connect segments belonging to the same tetrahedron. 
Notice that this partition function provides a quantization of YM theory only when the 
full series is considered. As discussed in \cite{BFYM}, the order-by-order equivalence requires 
additional gauge-fixing terms, which change the topological nature of the zeroth order and thus 
the spinfoam procedure used here.

The zeroth order of the coupled partition function \Ref{pert2} corresponds to the product
of two independent BF partition functions,
\equ\label{zero'th}
Z^{(0)}\equiv  Z[J,\eta]\Big|_{J=\eta=0}=Z_{\rm BF}[SU(2)] \,Z_{\rm BF}[{\cal G}].
\nequ
It is a sum over all the labelings of all the segments of the triangulation by irreps
of $\SU(2)$ and $\cal G$, where the $\SU(2)$ labels describe the geometry of the triangulation, as in the pure gravity picture, and the $\cal G$ labels describe the degrees of freedom of YM theory.
We report in the following table the homogeneity of the labeling,
which has been achieved discretizing both GR and YM connections on the dual triangulation:

\vspace{0.4cm}\begin{center}
\framebox{\begin{tabular}{l||c|c|c|c|c}
\emph{simplex} & \emph{GR labels} & \emph{YM labels} & \emph{GR amplitude}  & \emph{YM amplitude} & \emph{Interaction} \\
&&&&& \\
point 			& 			&	         &           &              		& \\
segment 		&		$j$	&  $\rho$  &	dim $j$  &	dim $\rho$      & \\
triangle    &		  	&	 $i$     & 	         & 								  & \\
tetrahedron &		    &          & \6	     & $A_\tau(\rho, i)$	  & 
$S_{\rm BB}[j_s, \rho_s, i_t]$
\end{tabular}}\end{center}
\vspace{0.4cm}
This structure allows us to identify 2d slices of the spinfoam
with a spin network state for the canonical quantization of the coupled system.
The links of such a spin network would be coloured by the labels for the irreps
of GR's $\SU(2)$ and YM's $\SU(N)$, and on the nodes by the intertwiners
of $\SU(N)$: $\ket{s} = \ket{\gamma, j_l, \rho_l, i_n}$.

In the next section we study the metric dependence of the interaction term $S_{\rm BB}[j_s, \rho_s, i_t]$.

\subsection{Effective action}\label{firstorder}
The first order term in the expansion \Ref{pert2} is
$Z^{(1)} \equiv -i\lambda \mean{S_{\rm BB}}$, thus it amounts to computing the effective
action for the interacting term \Ref{BwB}.
For the $\SU(2)$ graspings we use the diagrammatic notation \Ref{double1}, and we $T$-order them as
in \Ref{double3}, \Ref{triple3}. The double grasping on $\cal G$ has no ordering ambiguities (it appears
only linearly at this order, and it commutes with the other graspings).
Introducing the diagrammatic notation as in 
\Ref{double1}, we can write
\eqa \label{first2}
\mean{S_{\rm BB}} &=& \sum_{\{j_s, \rho_s, i_t\}} \int_0^\infty dt\;
\prod_s {\rm dim}j_s \; {\rm dim}\rho_s \;\sum_{\tilde\tau}\; \prod_{\tau\neq\tilde\tau}\;
\{6j\}\;A_\tau(\rho_s, i_t)
\Bigg[\Bigg(
\parbox{1.3cm}{\begin{picture}(0,0) (10,0)
\SetScale{0.12}\SetWidth{4}\SetColor{Black}\put(9,-3){$T\{$}
\DashLine(201,120)(201,-120){10}\DashLine(241,120)(241,-120){10}\Photon(331,120)(331,-120){8}{6}
\put(31,-3){$\}$}\end{picture}}
\Bigg|
\parbox{1.2cm}{\begin{picture}(0,0) (5,14)
\SetWidth{3}\SetScale{0.1}\SetColor{Black} 
\seij \Text(34,-4)[lb]{\small{$\tilde\tau$}}\end{picture}}\;\Bigg) + 
\no\no &&  -t\,
\Bigg(
\parbox{2.4cm}{\begin{picture}(0,0) (4,0)
\SetScale{0.12}\SetWidth{4}\SetColor{Black}\put(3,-3){$T\{$}
\DashLine(401,120)(401,-120){10}\DashLine(441,120)(441,-120){10}
\DashLine(241,120)(241,0){10}\DashLine(346,-105)(241,1){10}\DashLine(136,-105)(241,1){10}
\put(57,-3){$\}$} \Photon(541,120)(541,-120){8}{6} \end{picture}}
\Bigg|
\parbox{1.2cm}{\begin{picture}(0,0) (5,14)
\SetWidth{3}\SetScale{0.1}\SetColor{Black} 
\seij \Text(34,-4)[lb]{\small{$\tilde\tau$}}\end{picture}}\;\Bigg)+ \frac{t^2}{2}
\Bigg(
\parbox{3.4cm}{\begin{picture}(0,0) (-25,0)
\SetScale{0.12}\SetWidth{4}\SetColor{Black}\put(-25,-3){$T\{$}
\DashLine(401,120)(401,-120){10}\DashLine(441,120)(441,-120){10}
\DashLine(241,120)(241,0){10}\DashLine(346,-105)(241,1){10}\DashLine(136,-105)(241,1){10}
\DashLine(1,120)(1,0){10}\DashLine(106,-105)(1,1){10}\DashLine(-104,-105)(01,1){10}
\put(57,-3){$\}$}\Photon(545,120)(545,-120){8}{6}\end{picture}}
\Bigg|
\parbox{1.2cm}{\begin{picture}(0,0) (5,14)
\SetWidth{3}\SetScale{0.1}\SetColor{Black} 
\seij \Text(34,-4)[lb]{\small{$\tilde\tau$}}\end{picture}}\;\Bigg) + \ldots\Bigg]
 \neqa

\bigskip

The key diagram to evaluate is
\equ\label{meanC}
\mean{{\cal C}_\tau} =
\Bigg(\parbox{1.3cm}{\begin{picture}(0,0) (10,0)\SetScale{0.12}\SetWidth{4}\SetColor{Black}\put(9,-3){$T\{$}
\DashLine(201,120)(201,-120){10}\DashLine(241,120)(241,-120){10}\Photon(331,120)(331,-120){8}{6}
\put(31,-3){$\}$}\end{picture}} \Bigg|
\parbox{1.2cm}{\begin{picture}(0,0) (5,14)\SetWidth{3}\SetScale{0.1}\SetColor{Black} 
\seij \Text(34,-4)[lb]{\small{$\tau$}}\end{picture}}\;\Bigg).
\nequ
This diagram has $4\times 36=144$ contributions, coming from all the possible choices of graspings
in a given point, times the four points.
Each contribution can be evaluated using grasping rules and recoupling theory as in 
\cite{Hackett}. Because there are only double graspings entering this expression, the evaluation 
is rather simple and we do not report the details here, but only the asymptotics.

Let us distinguish two types of terms, when the YM grasping is diagonal,
namely $s_3=t_3$, and when is not diagonal, namely $s_3 \neq t_3$. 
Consider first the diagonal case.
For fixed $s_3=ij$, there are $4$ contributions from $p=i$, and four from $p=j$.
To fix ideas, let us choose $s_3=12$. The YM grasping immediately gives $C^2(\rho_{12})$
from \Ref{YMgrasps}. As for the GR grasping,
we can implement the antisymmetrization from the $\eps$ tensors of \Ref{Cgrasping}
in the algebraic indeces, and using \Ref{double3}
the large spin limit of the relevant graspings from $p=1$ gives
\equ\label{diag}
2 \, \mean{\, X_{13}^I \, X_{13}^{[I} \, X_{14}^J \, X_{14}^{J]} \ } 
\sim \f{\cos\left(S_{\rm R}(j_s)+\f\pi4\right)}{\sqrt{12\pi \, V(j_s)}} 
\, 2 \, \Big[\ell_{13}^2 \ell_{14}^2 - (\ell_{13}\cdot\ell_{14})^2\Big].
\nequ
The term in square brackets can be immediately recognized as $4 A_2{}^2$.
Analogously, the graspings from $p=2$ give $4 A_1{}^2$. There are in total 48 contributions
of this type.

The second case is when the YM grasping is non diagonal. Let us fix 
$s_3=12$, $t_3=13$. Choosing for simplicity $\SU(2)$, the YM grasping gives
$\f12[C^2(\rho_{23})-C^2(\rho_{12})-C^2(\rho_{13})]$. The GR
graspings have four contributions all coming from $p=1$, times 2 from the
symmetric choice $s_3=13$, $t_3=12$, giving in total the eight contributions
\equ\label{nondiag}
-\, 4 \, \mean{\, X_{13}^I \, X_{12}^{[I} \, X_{14}^J \, X_{14}^{J]} \ } 
\sim \f{\cos\left(S_{\rm R}(j_s)+\f\pi4\right)}{\sqrt{12\pi \, V(j_s)}} 
\, 4 \, \Big[(\ell_{12}\cdot\ell_{14}) (\ell_{13}\cdot\ell_{14}) - 
(\ell_{12}\cdot\ell_{13}) \ell_{14}^2 \Big].
\nequ
If we recall the definition of the dihedral angles $\theta_s$ in terms of angles $\phi_{ij}$ between the 
segment vectors (defined by $\ell_{ij}\cdot\ell_{ik} = \ell_{ij}\ell_{ik} \cos\phi_{jk}$), 
\equ
\sin\phi_{ij}\sin\phi_{jk} \cos\theta_{ik} = \cos\phi_{ij}\cos\phi_{jk}-\cos\phi_{ik},
\nequ
the term in square brackets in \Ref{nondiag} reads $\vec n_2 \cdot \vec n_3$.
Notice that the same contribution comes also from the choice $s_3=24$, $t_3=34$,
for which the YM grasping gives $\f12[C^2(\rho_{23})-C^2(\rho_{24})-C^2(\rho_{34})]$.
There are in total 96 contributions of this type. 

Adding up all the contributions we get 
\eqa\label{ciccio}
&& 8 \, C^2(\rho_{12}) \, (A_1^2 +A_2^2) + \ldots \no
&& - \, 4 \, \vec n_2 \cdot \vec n_3 \bigg[\f12\Big(C^2(\rho_{23})-C^2(\rho_{12})-C^2(\rho_{13}) \Big) +
\f12\Big(C^2(\rho_{23})-C^2(\rho_{24})-C^2(\rho_{34}) \Big) \bigg]+\ldots
\neqa
Collecting all the Casimirs and using the fact that $\sum_{i=1}^4 \vec n_i \equiv 0$, 
\Ref{ciccio} adds up to $-8\sum_{ij}  \vec n_i \cdot \vec n_j \, C^2(\rho_{ij})$. 
Therefore \Ref{meanC} has the following large spin behaviour,
\equ
\mean{{\cal C}_\tau} \sim - 2 \sum_{ij} \vec n_i \cdot \vec n_j \, C^2(\rho_{ij}) \, 
\f{\cos\left(S_{\rm R}(j_s)+\f\pi4\right)}{\sqrt{12\pi \, V(j_s)}}
\, \f{\cos\left(S_{\rm R}(\rho_s)+\f\pi4\right)}{\sqrt{12\pi \, V(\rho_s)}}
\nequ

Next, we look at the other diagrams entering \Ref{first2}. Thanks to the $T$-ordering, the 
powers of the triple grasping factorize in the large spin limit and one obtains simply 
\Ref{Vmen1}.
Putting everything together and approximating the sums over the irrep labels with integrals, we then have
\eqa\label{final}
-i  \, \lambda \, \mean{S_{\rm BB}} &\sim& \int \prod_s d{j_s} \, d \rho_s 
\prod_s {\rm dim}j_s \; {\rm dim}\rho_s \;\sum_{\tilde\tau} \prod_{\tau\neq\tilde\tau}\;
\{6j\}\;\{6\rho\} \times \no &\times&
\Bigg[\f{g_0{}^2}{12} \hbar {\lp} \sum_{ij} \f{\vec n_i \cdot \vec n_j}{V_{\tilde\tau}} \, C^2(\rho_{ij}) \Bigg]\, 
\f{\cos\left(S_{\rm R}(j_s)+\f34\pi\right)}{\sqrt{12\pi \, V(j_s)}}
\, \f{\cos\left(S_{\rm R}(\rho_s)+\f\pi4\right)}{\sqrt{12\pi \, V(\rho_s)}}.
\neqa
In the limit in which $C^2(\rho_{ij})\sim B_{ij}^a B_{ij}^a$, the term in square brackets is exactly 
$\f1\hbar S_{\rm RInt}$ given in \Ref{ReggeInt}. 
This is our key result: the large spin behaviour of the spinfoam model \Ref{pert2} is dominated
by the classical dynamics of GR coupled to YM described \`a la Regge.

\subsection{Tetrahedral asymptotics of the coupled partition function}
Let us now go back to the full partition function \Ref{pert2}, and consider a single tetrahedron. In this simple case, we can proceed as above to show the general result for powers of the operator $S_{\rm BB}$, 
\equ\label{general}
(-i \lambda )^n \, \mean{S_{\rm BB}{}^n} \sim \, 
(iS_{\rm RInt})^n \f{\cos\left(S_{\rm R}(\rho_s)+\f\pi4\right)}{\sqrt{12\pi\,V(j_s)} \, \sqrt{12\pi\,V(\rho_s)}}
\left\{ \begin{array}{lc}
i \, \cos\big(S_{\rm R}(j_s)+\f34\pi\big) & {\rm if} \ n \ {\rm is \ odd}, \\ \\
\cos\big(S_{\rm R}(j_s)+\f\pi4\big) & {\rm if} \ n \ {\rm is \ even}. 
\end{array}\right.
\nequ
Using \Ref{general} we can sum up the series \Ref{pert2} and evaluate the coupled partition function
in the large spin limit, 
\eqa\label{Zlimit}
Z &\sim& \int \prod_s d{j_s} \, d \rho_s \, \prod_s {\rm dim}j_s \; {\rm dim}\rho_s \
\f{\cos\left(S_{\rm R}(\rho_s)+\f\pi4\right)}{\sqrt{12\pi\,V(j_s)} \, \sqrt{12\pi\,V(\rho_s)}} \times \no\no\no
&\times& \Bigg[
\sum_{n=0}^\infty \f{(iS_{\rm RInt})^{2n+1}}{(2n+1)!} \, i \, 
\cos\left(S_{\rm R}(j_s)+\f34\pi\right) +
\sum_{n=0}^\infty \f{(iS_{\rm RInt})^{2n}}{(2n)!} \,   
\cos\left(S_{\rm R}(j_s)+\f\pi4\right)\Bigg] = \no\no\no
&=& \int \prod_s d{j_s} \, d \rho_s \, \prod_s {\rm dim}j_s \; {\rm dim}\rho_s \
\f{\cos\left( S_{\rm R}(j_s) - S_{\rm RInt}(j_s, \rho_s) + \f\pi4\right)\, 
\cos\left(   S_{\rm R}(\rho_s)+ \f\pi4\right)}
{\sqrt{12\pi\,V(j_s)} \, \sqrt{12\pi\,V(\rho_s)}}.
\neqa
We see from this result that the asymptotics of the tetrahedral partition function are dominated by 
linear combinations of exponentials of the classical discrete actions
described in section \ref{SecSemi}. Notice that in principle one would expect a single exponential
of the Regge version of \Ref{action4} from a correct semiclassical limit, namely
$$ e^{-i[S_{\rm R}(j_s) + S_{\rm R}(\rho_s) - S_{\rm RInt}(j_s, \rho_s)]}. $$
Yet the presence of the somewhat akward linear combination in \Ref{Zlimit} can be completely understood
if we recall that the spinfoam quantization includes a sum over both orientation of the (triangulated) spacetime manifold.
In fact, let us look at \Ref{action4}:
under change of orientation ($e_\mu^I\mapsto -e_\mu^I$) both the first and third terms change sign, but not the second.
This accounts for the first cosine term in \Ref{Zlimit},
$$ \left( e^{i[S_{\rm R}(j_s) - S_{\rm RInt}(j_s, \rho_s)]} 
+ e^{-i[S_{\rm R}(j_s) - S_{\rm RInt}(j_s, \rho_s)]}\right) e^{i S_{\rm R}(\rho_s)}. $$
The second cosine comes naturally when the same argument is applied also to the fibre manifold: under change
of orientation of the fibres ($B_\mu^a\mapsto -B_\mu^a$)
only the second term in \Ref{action4} changes sign, whereas the first and third do not.

So the tetrahedral partition function \Ref{Zlimit} is the correct semiclassical limit of a coupled quantum 
theory with the feature of summing over both orientations of the spacetime and fibre manifold.
However it is exactly the presence of this sum over orientations that makes 
problematic the interpretation of the partition function on a generic triangulation, whose semiclassical limit
we do not report here. In the PR model this is not a problem: the model is
topological invariant, thus the partition function on a generic triangulation can be reduced to the one on a single tetrahedron. The model presented here on the other hand is manifestly non topological invariant, 
thus this argument does not apply. 
We leave this question open, stressing that it is an issue of the PR model in its own, and not of the coupling
with YM performed here. Furthermore, notice that the problem can be circumvented by introducing
oriented boundary states in the computation of physical correlations, as it is done in the 
graviton calculations of \cite{grav}.

\section{Conclusions}
In this paper we defined a spinfoam model for riemannian 3d quantum gravity  coupled to Yang-Mills theory.
The partition function of this model is given in \Ref{pert2}. 
The construction makes use of the generating functional technique, and of grasping rules and 
$\SU(2)$ recoupling thery. With respect to previous attempts in the
literature, the model has the advantage of discretizing the gravitational and YM connections in the same 
way. This leads to  a homogeneous labeling of the triangulation by the GR and YM variables,
as reported in the table in section \ref{sectionPert}, and thus this model
provides transition amplitudes for spin networks of the
canonical theory. Furthermore, for the particular choice of $\SU(2)$ YM theory, we were able to show explicitly
that the model has the correct semiclassical limit, given by the discretization \`a la Regge of the
classical YM action. This result supports both the consistency of coupling matter fields to gravity in the way
proposed here, and the utility of the large spin limit to study semiclassical physics.
 
Let us add a few comments and perspectives on further developments.
One thing to stress is that as usual, the quantization procedure presents many ambiguities. In particular, here
we chose a definite way of discretizing the dynamical variables and the actions, and we gave a 
particular ordering prescription for the products of non commuting operators. Other choices
are possible. This one was preferred because it easily reproduced the correct
semiclassical limit. Once the correct limit is established, the key question becomes computing
the quantum gravity corrections to it. Already in the simple model presented here we see that this
is a formidable task due to the large number of contributions that the next to leading order of the asymptotics
receives. Yet this crucial question certainly deserves further work. 
Among the intriguing effects to look for there are modified dispersion relations: even though the
discrete structure of loop quantum gravity does not break Lorentz invariance per se \cite{Rovelli:2002vp},
it is commonly recognized that possible deformations of this symmetry is a promising sector
to look for observable quantum gravity effects \cite{Amelino-Camelia:2004hm}. Remarkably in 
\cite{PR3} it was shown that the coupling of point particles to the Ponzano-Regge model
gives a well defined non commutative effective action for the particles, 
thus resulting in a precise Lorentz deformation with modified dispersion relations.
It would be extremely interesting if analogue effects arise in the model here presented.\footnote{Notice
that the quantum
particle of 3d YM theory is massless and spinless, thus the results of \cite{PR3} might not apply straighforwardly.}
A potential difference lies in the fact that here the matter field is added to the gravitational
action, whereas in the approach of \cite{PR3} the particles are described using the 
gauge degrees of freedom of the gravitational sector. However the situation could be similar to what happens
in quantum field theory in the temporal gauge, for both YM theory \cite{Testa} and linearized gravity \cite{Mattei:2005cm},
where static matter sources (whose coupling is lost by the choice of gauge) turn out to be described by
gauge degrees of freedom.

At the end of the previous section we mentioned how an oriented boundary state could 
fix the orientation of the coupled partition function. Indeed, finding a suitable boundary state 
would have far more important applications: it would allow to follow the proposal developed
in \cite{grav} to compute scattering amplitudes for the coupled system described here. 
This would be a key development to understand low energy 
physics in the context of spinfoam gravity. From this point of view, the emergence of Regge calculus in the 
large spin limit is very promising, as this property is at the basis of the results in \cite{grav},
and can be used to construct a suitable boundary state, as discussed in \cite{Dittrich:2007wm}.

Another important question for future work is how to restore the
full diffeomorphism invariance of the continuum theory, here broken by the choice of
a fixed triangulation. Being the model non topological, the symmetry should be restored
by including a sum over the triangulations, for instance along the lines of 
the group field theory approach \cite{introgft}. 
Notice that restoring the right degrees of freedom can only be expected for the
fully resummed partition function. This is because as pointed out in section 6,
only the full partition function can be expected to be equivalent to YM theory.
This important aspect will be studied elsewhere.

Finally, it is sometimes argued that the spinfoam quantization of $\SU(2)$ BF theory simply
amounts to mapping the $B$ field to its eigenvalue $j$. This map has been considered in the literature,
especially in 4d, to define the spinfoam quantization of additional terms to the BF action, may they be constraints
reducing BF to GR, or matter fields. We would like to stress that particular care should
be taken in using this map: the more consistent procedure used in this paper, 
where the additional terms act as genuine quantum operators, shows that such a map would fail 
to capture the full quantum theory. In particular, one would not see the ordering ambiguities
and would fail to reproduce the full richness of the quantum corrections to the semiclassical leading order.

\section*{Acknowledgments}
The author is particularly grateful to Carlo Rovelli, Laurent Freidel, Hendryk Pfeiffer and John Barrett
for many discussions and suggestions.

Research at Perimeter Institute for Theoretical Physics is supported in
part by the Government of Canada through NSERC and by the Province of
Ontario through MRI.

\end{document}